\documentclass[%
 reprint,
 superscriptaddress,
 showpacs,preprintnumbers,
 amsmath,amssymb,
 aps,
 prl,
 longbibliography,
lengthcheck,%
]{revtex4-2}

\usepackage{graphicx}
\usepackage{dcolumn}
\usepackage{bm}
\usepackage{hyperref}
\usepackage{color}
\usepackage{ulem}

\begin{document}

\preprint{APS/123-QED}

\title{
Large photogalvanic spin current by magnetic resonance in bilayer Cr trihalides
}

\author{Hiroaki Ishizuka}
\affiliation{
Department of Physics, Tokyo Institute of Technology, Meguro, Tokyo, 152-8551, Japan
}

\author{Masahiro Sato}
\affiliation{
Department of Physics, Ibaraki University Mito, Ibaraki, 310-8512,  Japan
}

\date{\today}

\begin{abstract}
Magnetic materials show rich optical responses related to the magnetic order.
These phenomena reflect the nature of their excitations, providing a powerful probe for the magnetic states and a way to control them.
In recent years, such studies were extended to the optical control of spin current using nonlinear optical response similar to the photogalvanic effect.
However, neither a candidate material nor a general formula for calculating the photogalvanic spin current is known so far.
In this work, we develop a general theory for the photogalvanic spin current through a magnetic resonance process.
Using the nonlinear response formalism, we find the nonlinear conductivity consists of two contributions that involve one and two magnon bands;
the latter is a contribution unknown to date.
We argue that the two-band process produces a large photogalvanic spin current in the antiferromagnetic phase of bilayer CrI$_3$ and CrBr$_3$, whose resonance frequency can be tuned between GHz-THz range by an external magnetic field.
Our findings open a route to the studies on the photogalvanic effect of spin angular momentum in realistic setups.
\end{abstract}

\pacs{
}

\maketitle

\section{Introduction}

In a photogalvanic effect, a dc electric current occurs by the illumination of light~\cite{Sturman1992,Tan2016,Tokura2018}, such as in solar cells.
Phenomenologically, it is a nonlinear optical effect where the current $J_e$ reads $J_e=\sigma^{(2)}E(\omega)E(-\omega)$. Here, $E(\omega)$ is the intensity of oscillating electric field with the frequency $\omega$ and $\sigma^{(2)}$ is the nonlinear conductivity.
This phenomenon requires inversion symmetry breaking because both $J_e$ and $E(\omega)$ are odd under spatial inversion operation.
Recent studies revealed that the photogalvanic effects is a useful probe for non-trivial electronic states such as Weyl electrons~\cite{Ishizuka2016,Chan2017,deJuan2017,Ma2017,Osterhoudt2019,Rees2020} and Berry curvature dipole~\cite{Moore2010,Sodemann2015,Xu2018}.
Similar phenomena in magnetic excitations were also explored theoretically, 
where the photogalvanic spin current is generated by exciting one magnon~\cite{Proskurin2018,Bostrom2021}, a pair of magnons~\cite{Ishizuka2019a}, or by exciting spinons~\cite{Ishizuka2019b}.
For the magnetic excitations, the phenomenological formula reads $J_s=\sigma^{(2)}h(\omega)h(-\omega)$, where $J_s$ and $h(\omega)$ are the spin current and ac magnetic field with frequency $\omega$, respectively.
Similar to the electronic photogalvanic effect, the photogalvanic spin current also requires a noncentrosymmetric magnetic insulator.
Besides the inversion symmetry breaking, a material that shows a large spin current is favorable for the experimental investigation as the theoretical predictions so far~\cite{Ishizuka2019b,Ishizuka2019a} are small compared to the spin current by spin Seebeck effect~\cite{Uchida2010,Hirobe2017} and that by spin pumping~\cite{Kajiwara2010}.

\begin{figure}
  \includegraphics[width=\linewidth]{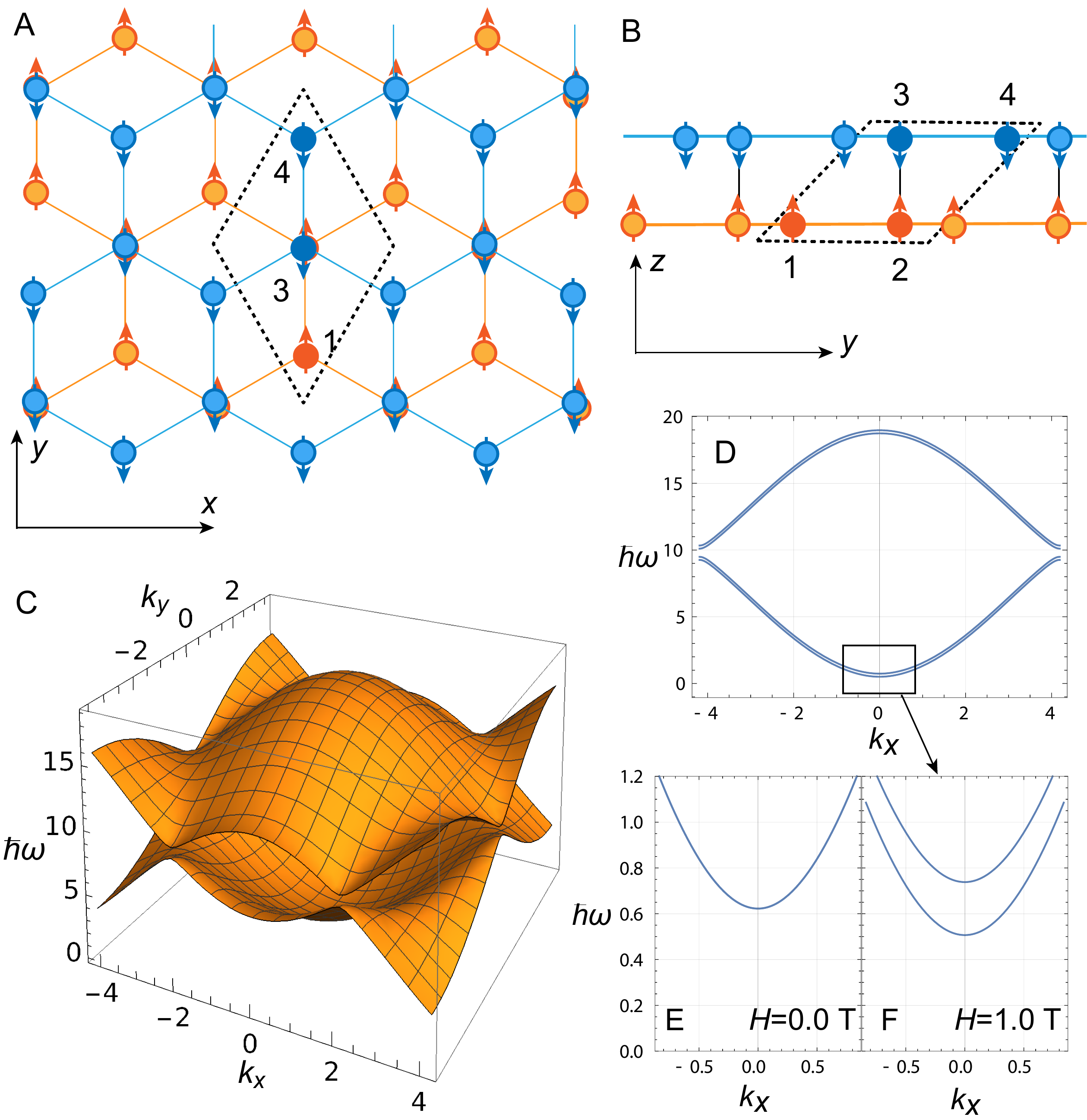}
  \caption{Schematic of the model and its magnon dispersion. (A) and (B) are the lattice structure and magnetic order of the antiferromagnetic Cr trihalides bilayer viewed from the $z$ and $x$ axes, respectively. The dashed diamond is the unit cell and the numbers 1-4 denote the sublattice indices in the unit cell. (C) Magnon band structure of CrI$_3$ in the first Brillouin zone. (D) Magnon dispersion along the $k_y=0$ line. (E) and (F) are the enlarged views of the lower magnon band around the $\Gamma$ point with $H=0$T (E) and $1$T (F). The parameters here are $J=2.01$ meV, $D_z=0.22$ meV, and $J_c=-0.59$ meV.}
  \label{fig:model}
\end{figure}

In view of the symmetry requirement, bilayer trihalides are an interesting candidate for studying the photogalvanic spin current.
Trihalide CrI$_3$ is a van der Waals magnet consisting of two-dimensional honeycomb layers of Cr $S=3/2$ spins (Fig.~\ref{fig:model}A).
A recent experiment discovered that a few layers of CrI$_3$ show magnetic orders at low temperatures, including the bilayer device~\cite{McGuire2015,Huang2017,Song2018}.
At low temperatures, the Cr spins in each honeycomb layer align ferromagnetically, forming a ferromagnetic sheet.
These ferromagnetic layers align antiferromagnetically under hydrostatic pressure~\cite{Li2019} or by applying electric field~\cite{Song2018,Jiang2019}.
Similar behavior is also known in CrBr$_3$, except that the magnetic anisotropy is weaker than CrI$_3$~\cite{Kim2019}.
In the paramagnetic phase, the bilayer CrI$_3$ has an inversion center at the middle of the two layers, whereas the antiferromagnetic order breaks the inversion symmetry~\cite{Zhang2019}. Hence, a photogalvanic spin current is allowed in the antiferromagnetic phase.

To study the photogalvanic spin current in material-specific models, we develop a general formula for the photogalvanic magnon spin current mediated by magnetic resonance. The formula is based on the nonlinear response theory, in which the resultant spin current conductivity consists of two contributions: the process only involves one magnon band, and the other involves two magnon bands.
The one-band contribution corresponds to the proposals in a previous study~\cite{Proskurin2018}, which vanishes in the model we consider.
The two-band process, on the other hand, is related to the off-diagonal component of spin-current operator giving a finite contribution in a system without DM interaction.
We find that the two-band contribution reach $\sigma^{(2)}\sim 10^{-10}$ Jcm$^{-2}$ in bilayer CrI$_3$ and CrBr$_3$ with Gilbert damping parameter $\alpha=10^{-2}$.
This estimate is orders of magnitude larger than other proposals~\cite{Ishizuka2019a,Ishizuka2019b}, predicting observable spin current density with $1$ mT ac magnetic field.
The conductivity is linearly proportional to $(\alpha\omega_0)^{-1}$ where $\omega_0$ is the resonance frequency, implying that $\sigma^{(2)}$ increases by reducing $\alpha$~\cite{Lenz2006,Vittoria2010} or by reducing $\omega_0$ by applying a magnetic field.
The photogalvanic spin current, if confirmed experimentally, should accelerate the study of nonlinear magnon transport in magnetic insulators and opto-spintronics~\cite{Nemec2018,Baltz2018}.

\section{Results}

\noindent
{\bf Spin model for Cr trihalides}\hspace{5mm} The effective spin model for bulk CrI$_3$ consists of layered honeycomb lattices of $S=3/2$ Cr spins.
The exchange interaction and anisotropy of the Cr spins are estimated from inelastic neutron-scattering experiment~\cite{Chen2018}, wherein they find a dominant intra-layer nearest-neighbor Heisenberg interaction and uniaxial anisotropy along with other small intra-layer interactions.
Hence, we consider an effective spin Hamiltonian with the nearest-neighbor ferromagnetic interaction $J$, interlayer antiferromagnetic interaction $J_c$, and the easy-axis anisotropy $D_z$.

The Hamiltonian reads
\begin{align}
H_0=
&-J\sum_{\substack{\langle in,jm\rangle}} \bm S_{in}\cdot\bm S_{jm}-J_c\sum_{i} \bm S_{i2}\cdot\bm S_{i3}\nonumber\\
&\qquad-D_z\sum_{in}(S^z_{in})^2-h\sum_{in}S^z_{in},\label{eq:H0}
\end{align}
where $\bm S_{in}\equiv(S^x_{in},S^y_{in},S^z_{in})$ is the $S=3/2$ Heisenberg spin on the sublattice $n,m=1,\cdots,4$ of $i$th unit cell.
Sublattices $n=1,2$ form the first honeycomb layer and $n=3,4$ sublattices form the second one. The final term is the Zeeman interaction with an external static field $h=g\mu_{\rm B}H$
($g$ is the g factor, $\mu_{\rm B}$ is the Bohr magneton, and $H$ is an applied static magnetic field along the $z$ axis).
The fitting of magnon bands to neutron scattering data gives $J=2.01$ meV, $D_z=0.22$ meV, and $J_c=0.59$ meV~\cite{Chen2018}.
The interlayer coupling in this estimate is ferromagnetic because the antiferromagnetic phase appears only by applying a gate voltage or by applying a pressure.
As the magnetic transition temperature in the antiferromagnetic phase is almost the same as that of the ferromagnetic phase, we take $J_c=-0.59$ meV.
With the antiferromagnetic $J_c$, the ground state of this model is an antiferromagnetic phase with the two ferromagnetic honeycomb layers align in an anti-parallel configuration (Fig.~\ref{fig:model}A).
Here, the spins point along the $z$ axis due to the uniaxial anisotropy $D_z$.
The effective Hamiltonian for CrBr$_3$ is similar to the CrI$_3$ Hamiltonian except for the values of the exchange interactions and the anisotropy~\cite{Samuelsen1971,Cai2021}, as we will discuss later; most importantly, the anisotropy is smaller in CrBr$_3$.

Figure~\ref{fig:model}C-\ref{fig:model}F shows the magnon band $\omega_{n\vec k}$ of CrI$_3$ using the above parameters.
Here, $\vec k=(k_x,k_y)$ is the wave vector of magnons and $n$ is the band index.
The magnon bands are doubly degenerate at $h=0$, while a finite field $h$ lifts the degeneracy due to the Zeeman splitting (Fig.~\ref{fig:model}F).
The calculated band structure is in semi-quantitative agreement with the recent observation~\cite{Cenker2021}.

We apply the ac transverse field
\begin{align}
H'=-h_x(t)\sum_{i,n}S^x_{in}-h_y(t)\sum_{i,n}S^y_{in}
\end{align}
to the system $H_0$, where $h_a(t)$ ($a=x,y$) are the magnetic field along $x$ and $y$ axes.
The nonlinear spin current conductivity for this perturbation is defined by
\begin{align}
	J^\alpha_\mu=\sum_{\mu,\nu}\int[\sigma^{(2)}]_{\mu\nu\lambda}^\alpha(0;\omega,-\omega)h_\nu(\omega)h_\lambda(-\omega) d\omega,
\end{align}
where $J^\alpha_\mu$ is the spin current for the $\alpha$ component of spin angular momentum flowing along the $\mu$ axis, and $h_\mu(\omega)=\int h_\mu(t)e^{-{\rm i}\omega t}dt$ is the Fourier transform of $h_\mu(t)$.
Here, only $J^z_\mu$ is discussed as $S^z$ is a conserved quantity, and hence, we can define the spin current unambiguously.
We derive the formula for $\sigma^{(2)}(0;\omega,-\omega)$ using a nonlinear response theory, which is similar to those for the photocurrent~\cite{Kraut1979} and two-magnon process~\cite{Ishizuka2019a} (see Method section for the formula and its derivation).

\begin{figure}
	\includegraphics[width=\linewidth]{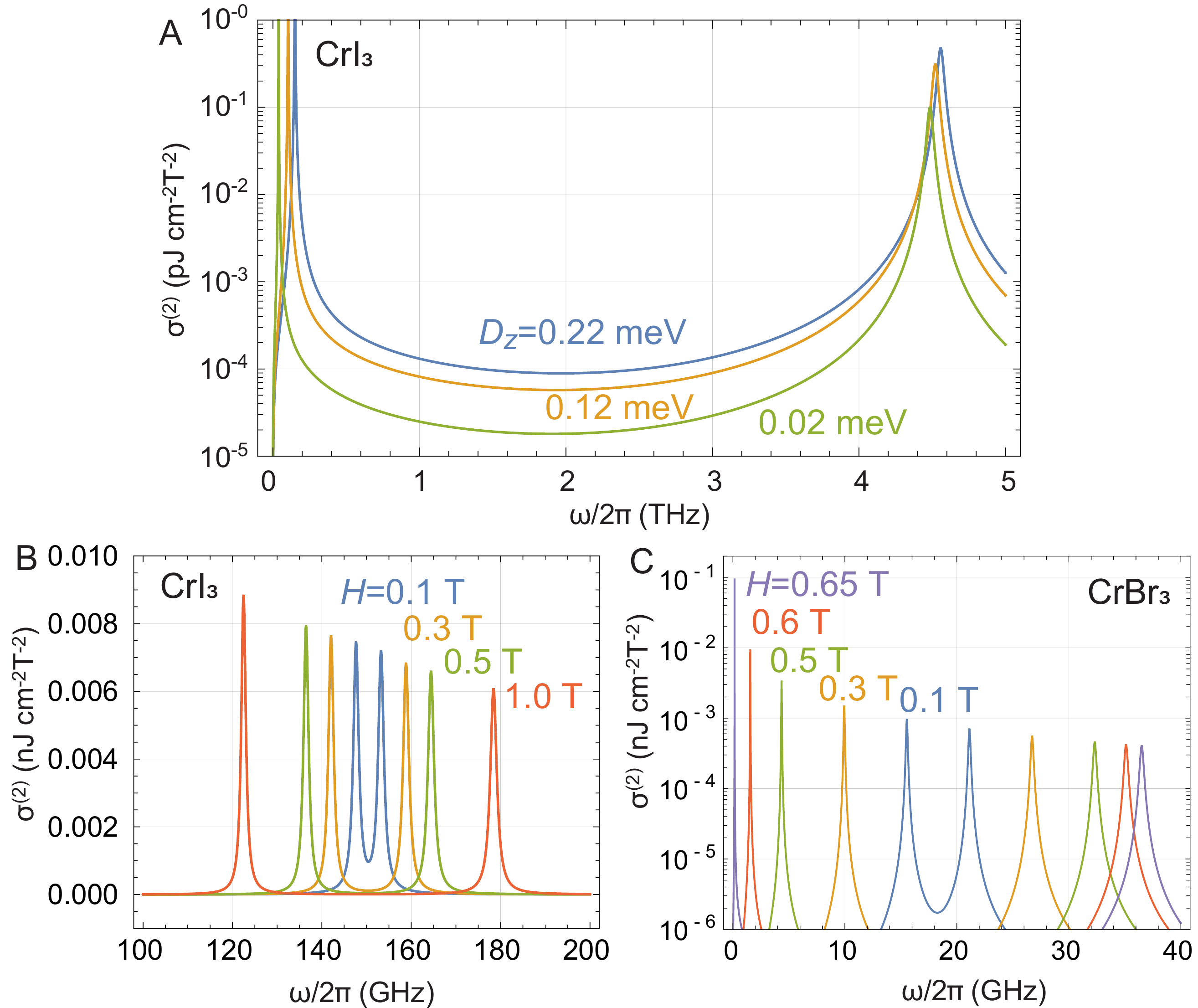}
	\caption{Spin current conductivity $[\sigma^{(2)}]^z_{yxx}(0;\omega,-\omega)$ for the model in Eq.~\eqref{eq:H0} with Gilbert damping $\alpha=10^{-2}$. (A) Frequency dependence of the spin current conductivity for different anisotropy $D_z=0.22$ meV (blue), $0.12$ meV (orange), and $0.02$ meV (green). The data for $D_z=0.22$ meV corresponds to CrI$_3$. Other parameters are $J=2.01$ meV, $J_c=-0.59$ meV, and $D_z=0.04$ meV. (B) Frequency dependence of the lower peaks for the static magnetic field $H=0.1$ T (blue), $0.3$ T (orange), $0.5$ T (green), and $1.0$ T (red).
	(C) Frequency dependence of the lower peaks for CrBr$_3$ with $H=0.1$ T, $0.3$ T, $0.5$ T, $0.6$ T, and $0.65$ T.
	The data are for $J=1.36$ meV, $J_c=-0.024$ meV, and $D_z=0.04$ meV.
	}
	\label{fig:Js}
\end{figure}

\noindent
{\bf Spin current conductivity}\hspace{5mm} Figure~\ref{fig:Js} shows the frequency $\omega$ dependence of $[\sigma^{(2)}]^z_{yxx}(0;\omega,-\omega)$ for the antiferromagnetic phase of the model in Eq.~\eqref{eq:H0}; the magnon relaxation rate reads $1/\tau=\alpha\omega$ where $\alpha=10^{-2}$ is the Gilbert damping constant.
We only show the results for $[\sigma^{(2)}]^z_{yxx}$ as $[\sigma^{(2)}]^z_{yyy}$ is the same as $[\sigma^{(2)}]^z_{yxx}$, and $[\sigma^{(2)}]^z_{xxy}$ is zero due to the symmetry of CrI$_3$ (See method section for details).
The position of two peaks in Fig.~\ref{fig:Js}A corresponds to the frequency of magnons at $\vec k=0$ in Fig.~\ref{fig:model}D.
The conductivity at the resonance peaks in Fig.~\ref{fig:Js} are $\sigma^{(2)}\sim10^{-11}$ Jcm$^{-2}$T$^{-2}$, and is linearly proportional to $\tau$.
Previous theories~\cite{Ishizuka2019a,Ishizuka2019b} argue that $J_s\sim 10^{-16}$ Jcm$^{-2}$ is necessary for the experimental observation of spin current.
According to Fig.~\ref{fig:Js}, $\sim1$ mT ac magnetic field ($\sim10^3-10^4$ Vcm$^{-1}$) is required to produce $J_s\sim 10^{-16}$ Jcm$^{-2}$ for $\tau=1/\alpha\omega_0=6.62\times10^{-10}$ s.
The required ac magnetic field is a couple of orders smaller than that of the mechanism in previous works~\cite{Ishizuka2019a,Ishizuka2019b}.

In Fig.~\ref{fig:Js}B, we show the magnetic field dependence of $[\sigma^{(2)}]^z_{y\nu\lambda}(0;\omega,-\omega)$.
The resonance frequency of the lower peak at zero static field is in the order of $10^2$ GHz, which is a consequence of the magnon gap induced by the Ising anisotropy of CrI$_3$.
Each peak split into two under the static magnetic field, reflecting the lifting of the degeneracy of magnon bands.
With increasing the magnetic field, the lower band eventually reaches zero energy causing a transition to a ferromagnetic phase.
The field-induced antiferromagnetic to ferromagnetic transition at $H\sim 0.5-1.0$~T is indeed observed in the experiment~\cite{Jiang2019,Li2019}.
Hence, the resonance frequency can be tuned by the external magnetic field within experimentally-available field strength.

Another route to tune the magnon gap is by changing the anisotropy.
A recent study reports that Cr trihalides with different halide ions have different anisotropy~\cite{Kim2019}:
CrI$_3$ is an easy-axis type magnet whereas CrBr$_3$ is almost Heisenberg-like with a small single-ion anisotropy, and the magnetic moments in CrCl$_3$ is XY like.
In addition, the interlayer coupling of CrBr$_3$ is controllable between ferromagnetic and antiferromagnetic by controlling the stacking~\cite{Chen2019}.
As in Fig.~\ref{fig:Js}A, the peak position for the antiferromagnetic order shifts to lower frequency as $D_z$ decreases.
In Fig.~\ref{fig:Js}C, we show the $H$ dependence of $[\sigma^{(2)}]_{yxx}^z(0;\omega,-\omega)$ for CrBr$_3$, in which case $J=1.36$ meV, $J_c=-0.024$ meV, and $D_z=0.04$ meV~\cite{Samuelsen1971,Cai2021}.
Here, we find that the peak $[\sigma^{(2)}]_{yxx}^z(0;\omega,-\omega)$ increases rapidly as the resonance frequency approaches zero; the result for $H=0.65T$ is two orders of magnitude larger than that of $H=0.1T$. Hence, as discussed below, reducing the resonance frequency by a magnetic field is a route to further enhance the spin current in a nearly isotropic material.

\noindent
{\bf Two-band process}\hspace{5mm}
We next turn to the mechanism of the photogalvanic response.
The nonlinear response formula reads
\begin{align}
	&[\sigma^{(2)}]_{\mu\nu\lambda}^\alpha(0;\omega,-\omega)=[\sigma^{(2;1b)}]_{\mu\nu\lambda}^\alpha(0;\omega,-\omega)\nonumber\\
	&\hspace{3cm}+[\sigma^{(2;2b)}]_{\mu\nu\lambda}^\alpha(0;\omega,-\omega),\label{eq:result:sigma}\\
&[\sigma^{(2;1b)}]_{\mu\nu\lambda}^\alpha(0;\omega,-\omega)=\nonumber\\
	&-\text{i}\frac\tau{\pi}\sum_{m=1}^{n_\text{uc}}\left[\frac{\tilde\beta^\nu_{m}[(\tilde J_{\vec0}^\mu)_{m,m}]\tilde\beta^\lambda_{m+n_\text{uc}}}{\omega-\omega_{m\vec0}-{\rm i}/2\tau}+\frac{\tilde\beta^\lambda_{m}[(\tilde J_{\vec0}^\mu)_{m,m}]\tilde\beta^\nu_{m+n_\text{uc}}}{\omega+\omega_{m\vec0}-{\rm i}/2\tau}\right],\nonumber
\end{align}
\begin{align}
	&[\sigma^{(2;2b)}]_{\mu\nu\lambda}^\alpha(0;\omega,-\omega)=-\frac1{2\pi}\sum_{\substack{m,l=1\\m\ne l}}^{n_\text{uc}}\frac1{\omega_{m\vec0}-\omega_{l\vec0}-{\rm i}/2\tau}\times\nonumber\\
	&\qquad\left[\frac{\tilde\beta^\nu_{l}[(\tilde J_{\vec0}^\mu)_{l,m}+(\tilde J_{\vec0}^\mu)_{m+n_\text{uc},l+n_\text{uc}}]\tilde\beta^\lambda_{m+n_\text{uc}}}{\omega-\omega_{l\vec0}-{\rm i}/2\tau}\right.\nonumber\\
	&\qquad\quad\left.+\frac{\tilde\beta^\lambda_{l}[(\tilde J_{\vec0}^\mu)_{l,m}+(\tilde J_{\vec0}^\mu)_{m+n_\text{uc},l+n_\text{uc}}]\tilde\beta^\nu_{m+n_\text{uc}}}{\omega+\omega_{m\vec0}-{\rm i}/2\tau}\right],\nonumber
\end{align}
where $\tilde J_{\vec0}^\mu$ is the $2n_\text{uc}\times2n_\text{uc}$ matrix of the spin current operator in the magnon eigenstate basis ($n_\text{uc}=4$ is the number of sublattices), $\omega_{m\vec k}$ is the eigen frequency of the $m$th magnon band with the momentum $\vec k$, and $\tilde\beta^\mu_{l}$ is the coupling constant between $h_\mu$ ($\mu=x,y$) and the $l$th $\vec k=\vec0$ magnon. The formal definition of $\tilde J_{\vec0}$ and $\tilde\beta^\mu_{l}$ is given in the Method along with the derivation of the formula.
The first term in Eq.~\eqref{eq:result:sigma}, $[\sigma^{(2;1b)}]_{\mu\nu\lambda}^z(0;\omega,-\omega)$, is the contrbution that involves one magnon band whereas the second term, $[\sigma^{(2;2b)}]_{\mu\nu\lambda}^z(0;\omega,-\omega)$, involves two magnon bands.

Unlike the one-band contribution $[\sigma^{(2;1b)}]_{\mu\nu\lambda}^z(0;\omega,-\omega)$, the two-band contribution $[\sigma^{(2;2b)}]_{\mu\nu\lambda}^z(0;\omega,-\omega)$ does not require Dzyaloshinskii-Moriya (DM) interaction.
The one-band contribution is proportional to the diagonal component of the spin current operator, hence, to the group velocity of magnons at $\vec k=\vec0$.
Therefore, DM interaction is necessary for the one-band process as studied in a previous work~\cite{Proskurin2018}.
In contrast, the two-band contribution is related to the off-diagonal components of the spin-current operator.
These terms generally remain nonzero at $\vec k=\vec0$ regardless of the symmetry of the band.
Therefore, the two-band process contributes to the photogalvanic spin current in a system with symmetric dispersion $\omega_{m\vec k}=\omega_{m,-\vec k}$, in contrast to the one-band process.

At the resonance frequency $\omega=\omega_{m\vec 0}$, the two band process is dominated by the resonating magnon band,
\begin{align}
&[\sigma^{(2;2b)}]_{\mu\nu\lambda}\sim-{\rm i}\frac{\tau}{\pi}\sum_{\substack{l=1\\l\ne m}}^{n_\text{uc}}\frac1{\omega_{m\vec0}-\omega_{l\vec0}-{\rm i}/2\tau}\times\nonumber\\
	&\qquad\left[\tilde\beta^\nu_{m}[(\tilde J_{\vec0}^\mu)_{m,l}+(\tilde J_{\vec0}^\mu)_{l+n_\text{uc},m+n_\text{uc}}]\tilde\beta^\lambda_{l+n_\text{uc}}\right.\nonumber\\
	&\qquad\quad\left.+\tilde\beta^\lambda_{l}[(\tilde J_{\vec0}^\mu)_{l,m}+(\tilde J_{\vec0}^\mu)_{m+n_\text{uc},l+n_\text{uc}}]\tilde\beta^\nu_{m+n_\text{uc}}\right].
\end{align}
The spin current is proportional to $\tau$, and hence, it is like the injection current in the photogalvanic effect.
This formula also implies that $[\sigma^{(2;2b)}]_{\mu\nu\lambda}$ increases linearly with $\tau=(\alpha\omega_{m\vec0})^{-1}$. Hence, a smaller resonance frequency is favorable as in Fig.~\ref{fig:Js}C.

\section{Discussion}

In this work, we developed a general theory for the photogalvanic spin current based on a nonlinear response theory.
Using the formula for the nonlinear conductivity, we predict that bilayer CrI$_3$ shows a large nonlinear spin current conductivity in the antiferromagnetic phase.
The conductivity shows sharp peaks at the frequency corresponding to the energy of $\vec k=\vec0$ magnon modes, resembling that of the magnetic resonance experiments.
The maximum spin current conductivity at the resonance frequency reaches $\sigma\sim10^{-10}$ Jcm$^{-2}$T$^{-2}$ for $\alpha=10^{-2}$, several orders of magnitude larger than that produced by other mechanisms~\cite{Ishizuka2019a,Ishizuka2019b}.
The estimated conductivity implies a spin current of $J_s\sim10^{-16}$ Jcm$^{-2}$ created by the application of $1$ mT ac magnetic field.
The spin current can be further enhanced by reducing the resonance frequency (Fig.~\ref{fig:Js}C), which is relevant to CrBr$_3$.
The required ac magnetic field for an observable spin current is a couple of orders smaller than those in the previous estimates~\cite{Ishizuka2019a,Ishizuka2019b}, hence, favorable for the experiment.

Recent studies on the few-layer Cr trihalides revealed that they are highly controllable two-dimensional magnets where both ferromagnetic and antiferromagnetic phases are realized by gating, hydrostatic pressure, and by applying external fields.
Our calculation shows that the photogalvanic spin current appears in the antiferromagnetic phase, whereas it is prohibited by symmetry in the ferromagnetic phase.
For the experiment, a setup similar to that in FMR studies should suffice~\cite{Lee2020,Zeisner2020}.
In addition, the direction of spin current changes depending on the orientation of the antiferromagnetic order, i.e., whether the magnetic moments on the first and second layers are up-down or down-up type.
These properties of the photogalvanic spin current gives an experimental identification for the photogalvanic spin current,
which should be detectable using the experimental setups discussed in Ref.~\cite{Ishizuka2019b}.

Optical technologies in the GHz to THz domain have experienced significant progress over recent years.
GHz waves have been long used in magnetic resonance experiments~\cite{Slichter1990}, and THz laser pulse techniques have been developed in the last decades~\cite{ Hirori2011,Cavalleri2017}.
In spintronics~\cite{Maekawa2017,Baltz2018}, such techniques are utilized to control magnetic states~\cite{Nemec2018}.
Intense-THz-laser driven phenomena in magnets have been also explored experimentally~\cite{Staub2014,Mukai2016,Nelson2017} and theoretically~\cite{Mochizuki2010,Sato2016,Sato2020,Kanega2021}.
The photogalvanic spin current proposed in this work should be detectable using the currently available techniques of GHz-THz waves.

\section{Method}

\subsection{Nonlinear response theory for free bosons}

We construct a general formula for the photogalvanic spin current by extending the linear response theory to the second-order in perturbation.
For the sake of generality, we here consider the spin current operator $J$ and the perturbation ${\cal H}'=\sum_\mu B^\mu F_\mu(t)$ where $B^\mu$ is a Hermitian operator and $F_\mu(t)$ is a time-dependent real field.
For the Zeeman coupling, $B^\mu=-\sum_i S_{i}^\mu$ ($\mu=x,y,z$) is the $\mu$ component of the total spin angular momentum and $F_\mu(t)=h_\mu(t)$ is the ac magnetic field ($S_{i}^\mu$ is the operator for the $\mu$ component of the spin on $i$th site).
We calculate the Fourier transform of the spin current within this setup, where the Fourier transform is defined by
\begin{align}
	J_\mu^\alpha(\Omega)=\int\langle {\cal J}_\mu^\alpha\rangle e^{-{\rm i}\Omega t} dt,
\end{align}
with $\langle {\cal J}_\mu^\alpha\rangle$ being the thermal average of spin current for $S^\alpha$ flowing along the $\mu$ axis and $\Omega$ being the frequency of the observed current.
By expanding the density matrix, as in the derivation of Kubo formula, the spin current reads
\begin{align}
J_\mu^\alpha(\Omega)=\sum_{\mu,\nu}\int [\sigma^{(2)}]_{\mu\nu\lambda}^\alpha(\Omega;\omega,\Omega-\omega) F_\nu(\omega)F_\lambda(\Omega-\omega) d\omega,
\end{align}
where
\begin{align}
&[\sigma^{(2)}]_{\mu\nu\lambda}^\alpha(\Omega;\omega,\Omega-\omega)=\frac1{2\pi}\sum_{n,m,l}\frac{(\rho_n-\rho_m)B_{nm}^\nu}{\omega+E_n-E_m-{\rm i}/2\tau}\times\nonumber\\
	&\quad\left(\frac{B_{ml}^\lambda [{\cal J}_\mu^\alpha]_{ln}}{\Omega+E_n-E_l-{\rm i}/2\tau}-\frac{[{\cal J}_\mu^\alpha]_{ml} B_{ln}^\lambda}{\Omega+E_l-E_m-{\rm i}/2\tau}\right),\label{eq:method:sigmagen}
\end{align}
is the nonlinear spin current conductivity. Here, $B_{nm}^\mu=\left<n\right| B^\mu\left|m\right>$ ($[{\cal J}_\mu^\alpha]_{nm}=\left<n\right|{\cal J}_\mu^\alpha\left|m\right>$) is the matrix elements for $B^\mu$ (${\cal J}^\alpha_\mu$) with $\left|m\right>$ being the ket vector of $m$th many-body states and $E_m$ being its eigenenergy. $\rho_n=e^{-\beta E_n}/Z$ is the statistical probability of the system in the $n$th many-body state, and $\tau$ is the magnon relaxation time.
Fourier transform of $F_\mu(t)$, $F_\mu(\omega)$,
is defined as
\begin{align}
F_\mu(\omega)=\int F_\mu(t)e^{-{\rm i}\omega t}dt,
\end{align}
where $\omega$ is the frequency of the applied ac field.
In general, calculating this equation for a spin model is a highly challenging task because the exact eigenstates are unknown in most cases.

To proceed further, we use the spinwave approximation focusing on the low-temperature limit.
Within the linear-spinwave approximation using Holstein-Primakov transformation, the ground state and the low-energy excitations of an ordered magnet are described by an effective boson Hamiltonian.
The general form of Hamiltonian for a magnetic ground state with $n_\text{uc}$ sublattice sites is
\begin{align}
{\cal H}_0=\sum_{\vec k}\psi^\dagger_{\vec k}{\cal H}_{\vec k}\psi_{\vec k},\label{eq:model:H}
\end{align}
where ${\cal H}_{\vec k}$ is a $2n_\text{uc}\times2n_\text{uc}$ Hermitian matrix and $\psi_{\vec k}={}^T(a_{1\vec k},\cdots,a_{n_\text{uc}\vec k},a_{1\vec k}^\dagger,\cdots,a_{n_\text{uc}\vec k}^\dagger)$ is the $2n_\text{uc}$ vector of magnon annihilation and creation operators.
Here, $a_{n\vec k}=1/\sqrt{N}\sum_{\vec R}a_{n}(\vec R)e^{\text{i}\vec k\cdot\vec R}$ and $a_{n}(\vec R)$ is the annihilation operator of a magnon on the $n$th sublattice of the unit cell at $\vec R$.
An example of the spinwave Hamiltonian is given in the main text for the Cr trihalides.
For the perturbation Hamiltonian, we assume
\begin{align}
B^\mu=\vec\beta^\mu\psi_{\vec 0},\label{eq:model:B}
\end{align}
where $\vec\beta^\mu=(\beta^\mu_1,\cdots,\beta^\mu_{2n_\text{uc}})$ is a $2n_\text{uc}$ component vector with $\beta^\mu_{i+n_\text{uc}}=(\beta^\mu_{i})^\ast$.
Similarly, we assume the spin current operator of form
\begin{align}
{\cal J}^\alpha_\mu=\sum_{\vec k}\psi_{\vec k}^\dagger {\cal J}^\alpha_{\mu,\vec k}\psi_{\vec k},\label{eq:model:J}
\end{align}
where ${\cal J}^\alpha_{\mu,\vec k}$ are $2n_\text{uc}\times2n_\text{uc}$ matrices. The argument from here on applies to arbitrary models written in the form of Eqs.~\eqref{eq:model:H}, \eqref{eq:model:B}, and \eqref{eq:model:J}.

The Hamiltonian in Eq.~\eqref{eq:model:H} is diagonalizable using a paraunitary matrix $T_{\vec k}$ that satisfies $T^{-1}_{\vec k}=\tilde IT_{\vec k}^\dagger \tilde I$~\cite{Corpa1978}.
Namely, there exists a matrix $T_{\vec k}$ such that $(T_{\vec k}^\dagger)^{-1}{\cal H}_{\vec k}T_{\vec k}^{-1}=\frac12 E_{\vec k}$ where $E_{\vec k}=\text{diag}(\omega_{1\vec k},\cdots,\omega_{n_\text{uc}\vec k},\omega_{1\vec k},\cdots,\omega_{n_\text{uc}\vec k})$ is a $2n_\text{uc}\times2n_\text{uc}$ diagonal matrix and $\omega_{n\vec k}$ is the eigenenergy of $n$th magnon band with the wavenumber $\vec k$.
Here, $\tilde I=\text{diag}(1,\cdots,1,-1,\cdots,-1)$ is the paraunit matrix.
A numerical algorithm for calculating $T$ for the general Hamiltonian is given in Ref.~\cite{Corpa1978}.
Using this method, without loss of generality, we can find a $T$ such that the eigenstate creation and annihilation operators are given by
\begin{align}
\alpha_{kn}=&\sum_{m=1}^{n_\text{uc}}T_{n,m}a_{kn}+T_{n,m+n_\text{uc}}a_{\vec km+n_\text{uc}}^\dagger,\nonumber\\
\alpha_{kn}^\dagger=&\sum_{m=1}^{n_\text{uc}}T_{n+n_\text{uc},m}a_{kn}+T_{n,m+n_\text{uc}}a_{\vec km+n_\text{uc}}^\dagger,
\end{align}
respectively. Introducing a vector of eigenstate operators $\phi_{\vec k}={}^T(\alpha_{k1},\cdots,\alpha_{kn_\text{uc}},\alpha_{k1}^\dagger,\cdots,\alpha_{kn_\text{uc}}^\dagger)$, Eqs.~\eqref{eq:model:H}, \eqref{eq:model:B}, and \eqref{eq:model:J} read
\begin{align}
&{\cal H}_0=\frac12\sum_{\vec k}\phi^\dagger_{\vec k}E_{\vec k}\phi_{\vec k},\quad
B^\mu=\vec{\tilde \beta}^\mu T^{-1}\phi_{\vec 0},\nonumber\\
&{\cal J}^\alpha_\mu=\sum_{\vec k}\phi_{\vec k}^\dagger\tilde{\cal J}^\alpha_{\mu,\vec k}\phi_{\vec k},\label{eq:method:spinwave}
\end{align}
respectively.
Here, $\vec{\tilde\beta}^\mu=\vec\beta^\mu T^{-1}$, and $\tilde{\cal J}^\alpha_{\mu,\vec k}=(T^\dagger)^{-1}{\cal J}^\alpha_{\mu,\vec k}T^{-1}$.
We here assume $(\tilde {\cal J}^\alpha_{\mu,\vec k})_{n+n_\text{uc},n+n_\text{uc}}=0$ for $1\le n\le n_\text{uc}$, namely, the spin current in the ground state is zero.
Equations~\eqref{eq:method:spinwave} show that the problem of the nonlinear response of spin systems reduces to that of free bosons.

At the zero temperature, the density matrix becomes $\rho_0=1$ and $\rho_n=0$ ($n>0$).
In this case, only the ground state, one-magnon states and two-magnon states with $\vec k=\vec0$ contributes to the sum.
Hence, the spin current conductivity formula in Eq.~\eqref{eq:method:sigmagen} reads
\begin{widetext}
\begin{align}
&[\sigma^{(2)}]_{\mu\nu\lambda}^\alpha(0;\omega,-\omega)=\nonumber\\
	&\qquad-\frac1{2\pi}\sum_{m,l=1}^{n_\text{uc}}\frac1{\omega_{m\vec0}-\omega_{l\vec0}-{\rm i}/2\tau}\left[\frac{\tilde\beta^\nu_{l}[(\tilde {\cal J}_{\mu,\vec0}^\alpha)_{l,m}+(\tilde {\cal J}_{\mu,\vec0}^\alpha)_{m+n_\text{uc},l+n_\text{uc}}]\tilde\beta^\lambda_{m+n_\text{uc}}}{\omega-\omega_{l\vec0}-{\rm i}/2\tau}+\frac{\tilde\beta^\lambda_{l}[(\tilde {\cal J}_{\mu,\vec0}^\alpha)_{l,m}+(\tilde {\cal J}_{\mu,\vec0}^\alpha)_{m+n_\text{uc},l+n_\text{uc}}]\tilde\beta^\nu_{m+n_\text{uc}}}{\omega+\omega_{m\vec0}-{\rm i}/2\tau}\right]\nonumber\\
	&\qquad+\frac1{2\pi}\sum_{m,l}\left[\frac1{\omega+\omega_{m\vec0}-{\rm i}/2\tau}\frac{[(\tilde {\cal J}_{\mu,\vec0}^\alpha)_{l+n_\text{uc},m}+(\tilde {\cal J}_{\mu,\vec0}^\alpha)_{m+n_\text{uc},l}](1+\delta_{ml})\beta^\lambda_{l+n_\text{uc}}\beta^\nu_{m+n_\text{uc}}}{\omega_{m\vec0}+\omega_{l\vec0}-{\rm i}/2\tau}\right.\nonumber\\
	&\hspace{5cm}\left.-\frac1{\omega-\omega_{m\vec0}-{\rm i}/2\tau}\frac{\beta^\nu_m\beta^\lambda_l(1+\delta_{ml})[(\tilde {\cal J}_{\mu,\vec0}^\alpha)_{m,l+n_\text{uc}}+(\tilde {\cal J}_{\mu,\vec0}^\alpha)_{l,m+n_\text{uc}}]}{\omega_{m\vec0}+\omega_{l\vec0}+{\rm i}/2\tau}\right],
\end{align}
\end{widetext}
where the sum over $m$ and $l$ are for magnon band index.
The first line in this formula is the one-magnon process that involves only one magnon states and the second and third lines are two magnon process involving two-magnon states.
The latter gives a finite contribution if the spin current operator has terms with two annihilation and creation operators, such as $\alpha_{m\vec0}\alpha_{m\vec0}$ and $\alpha_{m\vec0}^\dagger\alpha_{m\vec0}^\dagger$.
The cases we consider, however, do not have these terms.
In this case, the nonlinear spin-current conductivity reads
\begin{align}
&[\sigma^{(2)}]_{\mu\nu\lambda}^\alpha(0;\omega,-\omega)=-\frac1{2\pi}\sum_{m,l=1}^{n_\text{uc}}\frac1{\omega_{m\vec0}-\omega_{l\vec0}-{\rm i}/2\tau}\times\nonumber\\
	&\qquad\left[\frac{\tilde\beta^\nu_{l}[(\tilde {\cal J}_{\mu,\vec0}^\alpha)_{l,m}+(\tilde {\cal J}_{\mu,\vec0}^\alpha)_{m+n_\text{uc},l+n_\text{uc}}]\tilde\beta^\lambda_{l+n_\text{uc}}}{\omega-\omega_{l\vec0}-{\rm i}/2\tau}\right.\nonumber\\
	&\qquad\quad\left.+\frac{\tilde\beta^\lambda_{l}[(\tilde {\cal J}_{\mu,\vec0}^\alpha)_{l,m}+(\tilde {\cal J}_{\mu,\vec0}^\alpha)_{m+n_\text{uc},l+n_\text{uc}}]\tilde\beta^\nu_{m+n_\text{uc}}}{\omega+\omega_{m\vec0}-{\rm i}/2\tau}\right].
	\label{eq:method:sigma}
\end{align}
We used this formula to calculate the photogalvanic spin current in the main text.

\subsection{Spin wave theory for Cr trihalide}

We constructed the spin-wave Hamiltonian using Holstein-Primakov transformation.
Here, we focus on the antiferromagnetic phase with the spins in layer 1 (sublattices $n=1$ and $2$) points along the $z$ axis and the spins in layer 2 (sublattices $n=3$ and $4$) pointing anti-parallel to the $z$ axis.
The transformation for this antiferromagnetic order reads
\begin{align}
	S_{in}^z=&S-a_{in}^\dagger a_{in},\nonumber\\
	S_{in}^+=&\sqrt{2S}\left(1-\frac{a_{in}^\dagger a_{in}}{2S}\right)^{\frac12}a_{in},\nonumber\\
	S_{in}^-=&\sqrt{2S}a_{in}^\dagger\left(1-\frac{a_{in}^\dagger a_{in}}{2S}\right)^{\frac12},
\end{align}
for $n=1,2$ and
\begin{align}
	S_{in}^z=&a_{in}^\dagger a_{in}-S,\nonumber\\
	S_{in}^+=&\sqrt{2S}a_{in}^\dagger\left(1-\frac{a_{in}^\dagger a_{in}}{2S}\right)^{\frac12},\nonumber\\
	S_{in}^-=&\sqrt{2S}\left(1-\frac{a_{in}^\dagger a_{in}}{2S}\right)^{\frac12}a_{in},
\end{align}
for $n=3,4$.
Replacing the spin operators by the above transformation and leaving terms up to the order ${\cal O}(S^{\frac12})$,
the spin Hamiltonian transforms to
\begin{align}
H_0\sim&-JS\sum_{\langle in,jm\rangle}(a_{in}^\dagger a_{jm}+a_{jm}^\dagger a_{in})\nonumber\\
	&-J_cS\sum_{i} a_{i2}^\dagger a_{i2}+a_{i3}^\dagger a_{i3}+a_{i2}a_{i3}+a^\dagger_{i3}a^\dagger_{i2}\nonumber\\
	&+(3J+D)S\sum_{in}a_{in}^\dagger a_{in}\nonumber\\
	&+h\sum_n a_{i1}^\dagger a_{i1}+a_{i2}^\dagger a_{i2}-a_{i3}^\dagger a_{i3}-a_{i4}^\dagger a_{i4},\\
	=&\sum_{\vec k}[(3J+D)S+h](a_{\vec k1}^\dagger a_{\vec k1}+a_{\vec k4}^\dagger a_{\vec k4})\nonumber\\
	&\quad+[(3J+D-J_c)S+h](a_{\vec k2}^\dagger a_{\vec k2}+a_{\vec k3}^\dagger a_{\vec k3})\nonumber\\
	&\quad-J_cS(a_{\vec k2} a_{\vec k3}+a_{\vec k3}^\dagger a_{\vec k2}^\dagger)\nonumber\\
	&\quad+h_{\vec k} (a_{\vec k2}^\dagger a_{\vec k1}+a_{\vec k4}^\dagger a_{\vec k3})+h_{\vec k}^\ast(a_{\vec k1}^\dagger a_{\vec k2}+a_{\vec k3}^\dagger a_{\vec k4}).
\end{align}
where $a_{\vec kn}=(1/\sqrt{N})\sum_ia_{in}e^{-{\rm i}\vec k\cdot\vec r_i}$ and $h_{\vec k}=-JS(1+e^{{\rm i}\vec k\cdot(-1/2,-\sqrt3/2)a}+e^{{\rm i}\vec k\cdot(1/2,-\sqrt3/2)a})$.
The ac transverse field term reads
\begin{align}
	H'&=-h_x(t)\sqrt{\frac{S}2}\sum_{n=1}^4 a_{in}+a_{in}^\dagger\nonumber\\
	&+{\rm i}h_y(t)\sqrt{\frac{S}2}\left[\sum_{n=1}^2(a_{in}+a_{in}^\dagger)-\sum_{n=3}^4(a_{in}+a_{in}^\dagger)\right].
\end{align}
We used this model for the calculation in the main text.

\subsection{Spin current}

The definition of the spin current is often ambiguous as the spin angular momentum is not a conserved quantity.
However, in the model we consider, total $S^z$ is a conserved quantity.
Hence, we can define the spin current from the continuity equation, similar to the definition of electric current.
The operators for spin current density reads
\begin{align}
    J_\mu^z=\sum_{\langle in,jm\rangle} J[(r_j)_\mu-(r_i)_\mu](S_{in}^xS_{jm}^y-S_{in}^yS_{jm}^x).
\end{align}
Within the linear spin-wave theory, it reads
\begin{align}
    J_\mu^z=&\sum_{\langle i1,j2\rangle} {\rm i}JS[(r_{j2})_\mu-(r_{i1})_\mu](a_{i1}^\dagger a_{j2}-a_{j2}^\dagger a_{i1})\nonumber\\
&-\sum_{\langle i3,j4\rangle} {\rm i}JS[(r_{j4})_\mu-(r_{i3})_\mu](a_{i3}^\dagger a_{j4} -a_{j4}^\dagger a_{i3}),\nonumber\\
    =&-\sum_{\vec k} (\partial_{k_\mu}h_{\vec k}^\ast) a_{\vec k1}^\dagger a_{\vec k2}+(\partial_{k_\mu}h_{\vec k})a_{\vec k2}^\dagger a_{\vec k1}\nonumber\\
&+\sum_{\vec k} (\partial_{k_\mu}h_{\vec k}^\ast) a_{\vec k3}^\dagger a_{\vec k4}+(\partial_{k_\mu}h_{\vec k})a_{\vec k4}^\dagger a_{\vec k3}.
\end{align}

\subsection{Symmetry argument}

The second-order response is often prohibited by certain symmetries, as in the case of photocurrent.
We elaborate on the symmetry requirements for
a finite spin current, focusing on the inversion, mirror, and $\pi$-rotation operations.
The symmetry argument gives a rule for the direction of the photo-induced spin current discussed in the main text.

For concreteness, we focus on the spin current driven by ac magnetic field,
\begin{align}
J^\alpha_\mu=[\sigma^{(2)}]^\alpha_{\mu\nu\lambda}(0;\omega,-\omega)h^\nu(\omega) h^\lambda(-\omega).
\end{align}
For example, under the spatial inversion operation, the spin current and magnetic field transforms as $J_\mu^\alpha\to-J_\mu^\alpha$ and $h^\nu(\omega)\to h^\nu(\omega)$.
Hence, in a material with the inversion symmetry, we find
\begin{align}
[\sigma^{(2)}]^\alpha_{\mu\nu\lambda}(0;\omega,-\omega)=-[\sigma^{(2)}]^\alpha_{\mu\nu\lambda}(0;\omega,-\omega),
\end{align}
namely, the spin current conductivity vanishes in centrosymmetric materials.

\begin{table*}[tb]
    \centering
    \begin{minipage}{5.8cm}
    \begin{tabular}{c|ccccccccc}
    \hline
    $\sigma_x$ & $J_x^x$ & $J_x^y$  & $J_x^z$  & $J_y^x$  & $J_y^y$  & $J_y^z$ & $J_z^x$ & $J_z^y$ & $J_z^z$ \\
    \hline\hline
    $h_xh_x$   &-1       & 1        & 1        & 1        &-1        &-1       & 1       &-1       &-1 \\
    $h_xh_y$   & 1       &-1        &-1        &-1        & 1        & 1       &-1       & 1       & 1 \\
    $h_xh_z$   & 1       &-1        &-1        &-1        & 1        & 1       &-1       & 1       & 1 \\
    $h_yh_y$   &-1       & 1        & 1        & 1        &-1        &-1       & 1       &-1       &-1 \\
    $h_yh_z$   &-1       & 1        & 1        & 1        &-1        &-1       & 1       &-1       &-1 \\
    $h_zh_z$   &-1       & 1        & 1        & 1        &-1        &-1       & 1       &-1       &-1 \\
    \hline
    \end{tabular}
    \end{minipage}
    \begin{minipage}{6.0cm}
    \begin{tabular}{c|ccccccccc}
    \hline
    $\sigma_y$ & $J_x^x$ & $J_x^y$  & $J_x^z$  & $J_y^x$  & $J_y^y$  & $J_y^z$ & $J_z^x$ & $J_z^y$ & $J_z^z$ \\
    \hline\hline
    $h_xh_x$   &-1       & 1        &-1        & 1        &-1        & 1       &-1       & 1       &-1 \\
    $h_xh_y$   & 1       &-1        & 1        &-1        & 1        &-1       & 1       &-1       & 1 \\
    $h_xh_z$   &-1       & 1        &-1        & 1        &-1        & 1       &-1       & 1       &-1 \\
    $h_yh_y$   &-1       & 1        &-1        & 1        &-1        & 1       &-1       & 1       &-1 \\
    $h_yh_z$   & 1       &-1        & 1        &-1        & 1        &-1       & 1       &-1       & 1 \\
    $h_zh_z$   &-1       & 1        &-1        & 1        &-1        & 1       &-1       & 1       &-1 \\
    \hline
    \end{tabular}
    \end{minipage}
    \begin{minipage}{5.8cm}
    \begin{tabular}{c|ccccccccc}
    \hline
    $\sigma_z$ & $J_x^x$ & $J_x^y$  & $J_x^z$  & $J_y^x$  & $J_y^y$  & $J_y^z$ & $J_z^x$ & $J_z^y$ & $J_z^z$ \\
    \hline\hline
    $h_xh_x$   &-1       &-1        & 1        &-1        &-1        & 1       & 1       & 1       &-1 \\
    $h_xh_y$   &-1       &-1        & 1        &-1        &-1        & 1       & 1       & 1       &-1 \\
    $h_xh_z$   & 1       & 1        &-1        & 1        & 1        &-1       &-1       &-1       & 1 \\
    $h_yh_y$   &-1       &-1        & 1        &-1        &-1        & 1       & 1       & 1       &-1 \\
    $h_yh_z$   & 1       & 1        &-1        & 1        & 1        &-1       &-1       &-1       & 1 \\
    $h_zh_z$   &-1       &-1        & 1        &-1        &-1        & 1       & 1       & 1       &-1 \\
    \hline
    \end{tabular}
    \end{minipage}
    \caption{Table of symmetry-allowed current direction for $\sigma_a$ ($a=x,y,z$). The elements with $1$ indicates the conductivity for $J_\mu^\alpha$ generated by $h_\nu h_\lambda$ can be finite in presence of the symmetry $\sigma_a$, whereas that with $-1$ means the conductivity is zero due to the symmetry.}
    \label{tab:sigma}
\end{table*}

\begin{table*}[tb]
    \centering
    \begin{minipage}{5.8cm}
    \begin{tabular}{c|ccccccccc}
    \hline
    $C_2^x$ & $J_x^x$ & $J_x^y$  & $J_x^z$  & $J_y^x$  & $J_y^y$  & $J_y^z$ & $J_z^x$ & $J_z^y$ & $J_z^z$ \\
    \hline\hline
    $h_xh_x$   & 1       &-1        &-1        &-1        & 1        & 1       &-1       & 1       & 1 \\
    $h_xh_y$   &-1       & 1        & 1        & 1        &-1        &-1       & 1       &-1       &-1 \\
    $h_xh_z$   &-1       & 1        & 1        & 1        &-1        &-1       & 1       &-1       &-1 \\
    $h_yh_y$   & 1       &-1        &-1        &-1        & 1        & 1       &-1       & 1       & 1 \\
    $h_yh_z$   & 1       &-1        &-1        &-1        & 1        & 1       &-1       & 1       & 1 \\
    $h_zh_z$   & 1       &-1        &-1        &-1        & 1        & 1       &-1       & 1       & 1 \\
    \hline
    \end{tabular}
    \end{minipage}
    \begin{minipage}{6.0cm}
    \begin{tabular}{c|ccccccccc}
    \hline
    $C_2^y$ & $J_x^x$ & $J_x^y$  & $J_x^z$  & $J_y^x$  & $J_y^y$  & $J_y^z$ & $J_z^x$ & $J_z^y$ & $J_z^z$ \\
    \hline\hline
    $h_xh_x$   & 1       &-1        & 1        &-1        & 1        &-1       & 1       &-1       & 1 \\
    $h_xh_y$   &-1       & 1        &-1        & 1        &-1        & 1       &-1       & 1       &-1 \\
    $h_xh_z$   & 1       &-1        & 1        &-1        & 1        &-1       & 1       &-1       & 1 \\
    $h_yh_y$   & 1       &-1        & 1        &-1        & 1        &-1       & 1       &-1       & 1 \\
    $h_yh_z$   &-1       & 1        &-1        & 1        &-1        & 1       &-1       & 1       &-1 \\
    $h_zh_z$   & 1       &-1        & 1        &-1        & 1        &-1       & 1       &-1       & 1 \\
    \hline
    \end{tabular}
    \end{minipage}
    \begin{minipage}{5.8cm}
    \begin{tabular}{c|ccccccccc}
    \hline
    $C_2^z$ & $J_x^x$ & $J_x^y$  & $J_x^z$  & $J_y^x$  & $J_y^y$  & $J_y^z$ & $J_z^x$ & $J_z^y$ & $J_z^z$ \\
    \hline\hline
    $h_xh_x$   & 1       & 1        &-1        & 1        & 1        &-1       &-1       &-1       & 1 \\
    $h_xh_y$   & 1       & 1        &-1        & 1        & 1        &-1       &-1       &-1       & 1 \\
    $h_xh_z$   &-1       &-1        & 1        &-1        &-1        & 1       & 1       & 1       &-1 \\
    $h_yh_y$   & 1       & 1        &-1        & 1        & 1        &-1       &-1       &-1       & 1 \\
    $h_yh_z$   &-1       &-1        & 1        &-1        &-1        & 1       & 1       & 1       &-1 \\
    $h_zh_z$   & 1       & 1        &-1        & 1        & 1        &-1       &-1       &-1       & 1 \\
    \hline
    \end{tabular}
    \end{minipage}
    \caption{Table of symmetry-allowed current direction for $C^a_2$ ($a=x,y,z$). The elements with $1$ indicates the conductivity for $J_\mu^\alpha$ generated by $h_\nu h_\lambda$ can be finite in presence of the symmetry $C_2^a$, whereas that with $-1$ means the conductivity is zero due to the symmetry.}
    \label{tab:rotation}
\end{table*}

\begin{table}[tb]
    \centering
    \begin{minipage}{5.8cm}
    \begin{tabular}{c|ccccccccc}
    \hline
    $C_2^yC_2^z$ & $J_x^x$ & $J_x^y$  & $J_x^z$  & $J_y^x$  & $J_y^y$  & $J_y^z$ & $J_z^x$ & $J_z^y$ & $J_z^z$ \\
    \hline\hline
    $h_xh_x$   & 1       &-1        &-1        &-1        & 1        & 1       &-1       & 1       & 1 \\
    $h_xh_y$   &-1       & 1        & 1        & 1        &-1        &-1       & 1       &-1       &-1 \\
    $h_xh_z$   &-1       & 1        & 1        & 1        &-1        &-1       & 1       &-1       &-1 \\
    $h_yh_y$   & 1       &-1        &-1        &-1        & 1        & 1       &-1       & 1       & 1 \\
    $h_yh_z$   & 1       &-1        &-1        &-1        & 1        & 1       &-1       & 1       & 1 \\
    $h_zh_z$   & 1       &-1        &-1        &-1        & 1        & 1       &-1       & 1       & 1 \\
    \hline
    \end{tabular}
    \end{minipage}
    \caption{Table of symmetry-allowed current direction for $C^a_2$ ($a=x,y,z$). The elements with $1$ indicates the conductivity for $J_\mu^\alpha$ generated by $h_\nu h_\lambda$ can be finite in presence of the symmetry $C_2^yC_2^z$, whereas that with $-1$ means the conductivity is zero due to the symmetry.}
    \label{tab:CxCy}
\end{table}

We can perform similar analyses for mirror operations $\sigma_x$ ($x\to-x,y\to y,z\to z$), $\sigma_y$,  $\sigma_z$,  $C_2^x$ ($x\to  x,y\to-y,z\to-z$), $C_2^y$, and $C_2^z$.
When these symmetry exists, some components of $[\sigma^{(2)}]^\alpha_{\mu\nu\lambda}(0;\omega,-\omega)$ are prohibited by these symmetries.
Tables~\ref{tab:sigma} and \ref{tab:rotation} summarize the results of symmetry analysis.
For instance, -1 in the $J_x^x$ column and $h_xh_x$ row of $\sigma_x$ table means the nonlinear spin current $J_x^x$ by $h_x(\omega)h_x(-\omega)$ is prohibited in a material with $\sigma_x$.
On the other hand, 1 in the $J_x^x$ column and $h_xh_x$ row of $C_2^x$ table means the nonlinear spin current $J_x^x$ by $h_x(\omega)h_x(-\omega)$ is {\it not} prohibited by $C_2^x$.
It is, however, possible that other symmetries in the material prohibit the spin current.

We can construct the table for the products of two symmetry operations.
For example, when $C_2^yC_2^z$ exists, the table for $C_2^yC_2^z$ is obtained by taking the product of the same element in the tables for $C_2^y$ and $C_2^z$.
It is shown in Tab.~\ref{tab:CxCy}.
Similarly, one can straightforwardly construct the table for other combinations of symmetry operations as well.

For the case of two-dimensional magnets, we are particularly interested in $J_x^z$ and $J_y^z$, i.e., the spin current for $S^z$ that flows in the $xy$ plane.
We also restrict our selves to the transverse fields $h_x(\omega) h_x(-\omega)$ and $h_y(\omega) h_y(-\omega)$ as we are interested in the single-magnon processes.
In the effective model for CrI$_3$ and CrBr$_3$, the inversion symmetry prohibits the spin current in bulk and the single layer.
In contrast, $C_2^yC_2^z$ in the antiferromagnetic phase of the bilayer trihalides prohibit $J_x^z$ by $h_x(\omega) h_y(-\omega)$ and $h_y(\omega)h_y(-\omega)$ whereas $J_y^z$ is allowed.
Hence, we focus on $J_y^z$ in the main text.

\section{Acknowledgement}

This work is supported by JSPS KAKENHI (Grant Numbers JP18H03676, JP19K14649, and JP20H01830) and a Grant-in-Aid for Scientific Research on Innovative Areas "Quantum Liquid Crystals" (Grant No. JP19H05825).


%
\end{document}